\documentstyle [cite,rotating,epsf,twoside,11pt,here]{article}

\def\pt  {$p_{T}$}
\def\xf  {$x_{F}$}

\def\asymd  {$A(\mathrm{D^{\pm}})$}
\def\asymds {$A(\mathrm{D}_{s})$}
\def\asymlc {$A(\mathrm{\Lambda_{c}})$}

\def\pimin {$\pi^{-}$}

\def\sigmamin {$\Sigma^{-}$}

\def\lambdazero {$\Lambda^{0}$}

\def\Dz  {D$^{0}$}
\def\Daz {$\overline{\mathrm{D}}^{0}$}
\def\Dm  {D$^{-}$}
\def\Dp  {D$^{+}$}
\def\Dpm {D$^{\pm}$}
\def\Dsm {D$_{\mathrm{s}}^{-}$}
\def\Dsp {D$_{\mathrm{s}}^{+}$}
\def\Ds  {D$_{\mathrm{s}}$}

\def\Lc  {$\Lambda_{\mathrm{c}}$}
\def\Lca {$\overline{\Lambda}_{\mathrm{c}}$}
\def\Pccb {$c\overline{c}$}

\def\kevc1{\ifmmode\mathrm{\ keV/{\mit c}}
          \else$\mathrm{\ keV/{\mit c}}$\fi}
\def\Mevc1{\ifmmode\mathrm{\ MeV/{\mit c}}
          \else$\mathrm{\ MeV/{\mit c}}$\fi}
\def\gevc1{\ifmmode\mathrm{ GeV/{\mit c}}
          \else$\mathrm{ GeV/{\mit c}}$\fi}
\def\kevc2{\ifmmode\mathrm{\ keV/{\mit c}^2}
          \else$\mathrm{\ keV/{\mit c}^2}$\fi}
\def\Mevc2{\ifmmode\mathrm{\ MeV/{\mit c}^2}
          \else$\mathrm{\ MeV/{\mit c}^2}$\fi}
\def\Gevc2{\ifmmode\mathrm{\ GeV/{\mit c}^2}
          \else$\mathrm{\ GeV/{\mit c}^2}$\fi}
\def\Gev2c2{\ifmmode\mathrm{\ GeV^2/{\mit c}^2}
          \else$\mathrm{\ GeV^2/{\mit c}^2}$\fi}

\def\Pp{\ifmmode{\mathrm p}
         \else${\mathrm p}$\fi}
\def\Pn{\ifmmode{\mathrm n}
         \else${\mathrm n}$\fi}

\def\pimin {$\pi^{-}$}

\def\sigmamin {$\Sigma^{-}$}

\def\lambdazero {$\Lambda^{0}$}

\def\pt  {$p_{T}$}
\def\xf  {$x_{F}$}

\begin{document}

\pagestyle{empty}

\begin{flushright}
CERN-EP/98-41 \\
MPI-K H V5 1998 \\*[8mm]
6 March 1998
\end{flushright}
\vspace{0.8cm}
\Large
           
\Large
\centerline{
\parbox{12cm}{\begin{center}
Charge Asymmetries for D, \Ds{} and \Lc{} Production
in \mbox{\sigmamin{}~ - Nucleus} Interactions at 340 \gevc1{}
\end{center}
}}

\large
\vspace{1.8cm}
\centerline{The WA89 Collaboration}
\vspace{0.4cm}
\normalsize

\noindent
\sloppy
    M.I.~Adamovich$^8$,
  Yu.A.~Alexandrov$^8$,
       D.~Barberis$^3$,
           M.~Beck$^5$,
        C.~B\'erat$^4$,
         W.~Beusch$^2$,
           M.~Boss$^6$,
          S.~Brons$^{5,a}$,
     W.~Br\"uckner$^5$,
      M.~Bu\'enerd$^4$,
          C.~Busch$^6$,
      C.~B\"uscher$^5$,
      F.~Charignon$^4$,
        J.~Chauvin$^4$,
     E.A.~Chudakov$^{6,b}$,
         U.~Dersch$^5$,
       F.~Dropmann$^5$,
     J.~Engelfried$^{6,c}$,
         F.~Faller$^{6,d}$,
       A.~Fournier$^4$,
   S.G.~Gerassimov${^5,8}$,
      M.~Godbersen$^5$,
    P.~Grafstr\"om$^2$,
        Th.~Haller$^5$,
       M.~Heidrich$^5$,
        E.~Hubbard$^5$,
        R.B.~Hurst$^3$,
   K.~K\"onigsmann$^{5,e}$,
        I.~Konorov$^{5,8}$,
         N.~Keller$^6$,
        K.~Martens$^{6,f}$,
        Ph.~Martin$^4$,
     S.~Masciocchi$^{5,g}$,
       R.~Michaels$^{5,h}$,
       U.~M\"uller$^7$,
           H.~Neeb$^5$,
        D.~Newbold$^1$,
            C.~Newsom$^i$,
           S.~Paul$^{5,j}$,
    J.~Pochodzalla$^5$,
   I.~Potashnikova$^5$,
           B.~Povh$^5$,
        R.~Ransome$^k$,
            Z.~Ren$^5$,
  M.~Rey-Campagnolle$^{4,l}$,
         G.~Rosner$^7$,
          L.~Rossi$^3$,
        H.~Rudolph$^7$,
         C.~Scheel$^m$,
        L.~Schmitt$^{7,j}$,
     H.-W.~Siebert$^6$,
          A.~Simon$^{6,e}$,
        V.J.~Smith$^{1,n}$,
       O.~Thilmann$^6$,
       A.~Trombini$^5$,
          E.~Vesin$^4$,
       B.~Volkemer$^7$,
      K.~Vorwalter$^5$,
       Th.~Walcher$^7$,
       G.~W\"alder$^6$,
        R.~Werding$^5$,
       E.~Wittmann$^5$, and
   M.V.~Zavertyaev$^8$

\vspace{0.2cm}
\begin{flushleft}
\noindent
$^1${\sl University of Bristol, Bristol BS8 1TL, United Kingdom.}\\
$^2${\sl CERN, CH-1211 Gen\`eve 23, Switzerland.}\\
$^3${\sl Dipartimento di Fisica and I.N.F.N., Sezione di Genova,
        I-16146 Genova, Italy.}\\
$^4${\sl ISN, CNRS/IN2P3-UJF, 53 Avenue des Martyrs,
        F-38026 Grenoble Cedex, France.}\\
$^5${\sl Max-Planck-Institut f\"ur Kernphysik Heidelberg,
        D-69029 Heidelberg, Germany.}\\
$^6${\sl Physikalisches Institut, Universit\"at Heidelberg,
        D-69120 Heidelberg, Germany.$^o$}\\
$^7${\sl Institut f\"ur Kernphysik, Universit\"at Mainz,
        D-55099 Mainz, Germany.$^o$}\\
$^8${\sl Moscow Lebedev Physics Institute,
        RU-117924 Moscow, Russia.}\\
\end{flushleft}
\vspace*{1.0cm}

\normalsize
\centerline{\underline{\large{Abstract}}}
\vspace{0.5cm}

CERN experiment WA89 has studied charmed particles produced by a \sigmamin{} beam
at 340\gevc1{} on nuclear targets.  
Production of particles which have light quarks in common with
the beam is compared to production of those which do not.
Considerable production asymmetries between 
   \Dm{} and \Dp{},
   \Dsm{} and \Dsp{} and
   \Lc{} and \Lca{}
are observed. The results are compared with pion beam data and with theoretical
calculations.

\vspace*{1.0cm}
\centerline{(Submitted to European Physical Journal C)}

\newpage
\noindent
    $^a${\sl Now at TRIUMF, Vancouver, B.C., Canada V6T 2A3.}\\
    $^b${\sl On leave of absence from Moscow State University, 119889 Moscow, Russia.} \\
    $^c${\sl Now at FNAL, PO Box 500, Batavia,
          IL 60510, USA.} \\
    $^d${\sl Now at Fraunhofer Institut f\"ur Solarenergiesysteme,
               D-79100 Freiburg, Germany.}\\
    $^e${\sl Now at Fakult\"at f\"ur Physik, Universit\"at Freiburg, Germany.}\\
    $^f${\sl Now at Institute for Cosmic Ray Research, University of
             Tokyo,Japan.} \\
    $^g${\sl Now at Max-Planck-Institut f\"ur Physik, M\"unchen,
      Germany}\\
    $^h${Now at Thomas Jefferson Lab, Newport News, VA 23606, USA.} \\
    $^i${\sl University of Iowa, Iowa City, IA 52242, USA.}\\
    $^j${\sl Now at Technische Universita\"at M\"unchen, Garching, Germany}\\
    $^k${\sl Rutgers University, Piscataway, NJ 08854, USA.}\\
    $^l${\sl permanent address: CERN, CH-1211 Gen\`eve 23, Switzerland}\\
    $^m${\sl NIKHEF,1009 DB Amsterdam, The Netherlands.}\\
    $^n${\sl supported by the UK PPARC.}\\
    $^o${\sl supported by the Bundesministerium f\"ur Bildung, Wissenschaft,
          Forschung und Technologie, Germany, under contract numbers
          05~5HD15I, 06~HD524I and 06~MZ5265.}\\
\newpage

\pagestyle{plain}

\hyphenation{spec-tro-me-ter}
\hyphenation{charm-ed}

\setcounter{page}{1}
\section{Introduction}
\label{sec_1}

In hadronic interactions, typically,
the forward production of particles which have valence quarks/antiquarks in common
with the beam (leading particles)
is enhanced 
over those which do not
(non--leading particles).
This ``leading particle effect'' is strongly pronounced and well known in the light
quark sector, for example in the production of strange particles (see for example
\cite{WA89HYP} and the references therein).
Hadroproduction of charmed particles is usually described by splitting
it into a ``hard'' process of \Pccb{} production and a ``soft'' process
of hadronisation of $c$/$\overline{c}$ quarks to real hadrons. 
The production of \Pccb{} pairs 
has been described by next-to-leading order (NLO) PQCD calculations
\cite{NASO89, FRIX94} which do not predict significant asymmetries
between the $c$ and $\overline{c}$ produced \cite{E79196}.
For the hadronisation, however, only phenomenological models exist. 
Since no charm/anti--charm production asymmetries have been observed in
$e^{+}e^{-}$ interactions, while large asymmetries have been observed 
in charm hadroproduction, the observed asymmetries
can be accounted for by recombination of charm quarks/anti--quarks with 
the valence quarks from the beam particle.
In this picture the leading particle effect 
would be expected to increase at larger \xf{}, which indeed has been observed
(see later).
Since the differential cross--sections of charmed
particles are observed to drop sharply with increasing \xf{}, 
no considerable asymmetry between cross--sections of leading and non--leading 
particles integrated for \xf{}$>$0~ is expected. 

At present, there are high statistics data on 
\Dm{} and \Dp{} production in \pimin{} beams 
\cite{WA8293,E76994,E79196,WA9297},
which show a considerable enhancement of leading \Dm{} over 
non-leading \Dp{} in the kinematic range of \xf{}$>$0.3~. 
The asymmetry parameter A:
\begin{equation}
      A(x_{F}) = 
      \frac{d^{3}\sigma{}_{L}/dp^{3}-d^{3}\sigma{}_{NL}/dp^{3}}
           {d^{3}\sigma{}_{L}/dp^{3}+d^{3}\sigma{}_{NL}/dp^{3}}
\end{equation}
constructed using differential cross--sections of the leading and non--leading
particles,
rises with \xf{} and reaches values of A$\approx{}$0.5 at \xf{}=0.6~.
No asymmetry between \Dsm{} and \Dsp{} or \Lc{} and \Lca{} production 
by pion beams is observed \cite{WA9297,E79197,ACCM90}. 
In contrast to the high statistics
of the pion beam data (up to 75000 reconstructed \Dm{} plus \Dp{} in
\cite{E79196}) 
only limited data exist on charmed particle production by 
other hadron beams.
The results obtained 
in pp--interactions at 400~GeV/c 
by the the LEBC-EHS experiment 
\cite{EHSP88} which has observed about 50 \Dm{} plus \Dp{} candidates
indicated a ``harder'' \xf{} spectrum for
the non--leading \Dp{} than for the leading 
\Dm{}. 
No significant difference in the number of observed
\Lc{} and \Lca{} was found in a sample of 7 candidates.
Due to the low statistics no distinction between leading and non--leading
particles produced by 200~GeV/c protons in the ACCMOR experiment\cite{ACCM88} was done. 
In a proton beam of 250~GeV/c of experiment E769\cite{E769a96} 
an asymmetry integrated for \xf{}$>$0
of \asymd{}=0.18$\pm{}0.05$ has been observed, comparable to the
asymmetry measured in the pion beam of the same experiment \cite{E76994}
of \asymd{}=0.18$\pm{}0.06$.
\footnote{It is not quite clear from the paper\cite{E769a96} why the errors
          for these two asymmetries happen to be the same, while 
          one may infer from the paper that the statistics obtained in
          the pion beam was about 4 times higher than 
          in the proton beam.
         } 
The same experiment observed 
a large asymmetry \asymlc{}$>0.6 $ with 
a limited sample of about 35~\Lc{}.
These results indicate a leading particle effect in charm production
by protons, in contrast to the older results from \cite{EHSP88}.
In kaon beams it is of a particular interest to compare the production
of \Dsp{} and \Dsm{}.
The only asymmetry result comes from the same experiment \cite{E769a96}
which measured \asymds{} integrated for \xf{}$>$0   
and reported a value of 0.25$\pm{}$0.11.

It is therefore interesting to look for the leading particle
effect in baryon beams as well as in strange beams.
The most recent hyperon beam experiment, WA89 at CERN,
used a \sigmamin{} beam which
offers the opportunity to look for leading particle effects with
charmed--strange particles and with charmed baryons.
In this paper the results on production of \Dm{} versus \Dp{},
\Dsm{} versus \Dsp{} and \Lc{} versus \Lca{} are presented.
The first particle in each pair is leading in a \sigmamin{} beam 
while the second particle is not.
Neutral \Dz{} and \Daz{} are not considered since  
the direct production is non--leading for both particles while a considerable 
fraction of them can come from decays of charged higher states.
\\
%
\section{Experimental Setup}
\label{sec_setup}
%

Experiment WA89 was performed using the charged hyperon beam\cite{WA89BEAM} 
of the CERN SPS. The setup was based on a magnetic spectrometer
which detected decays of charmed particles 
produced at \xf{}$>$0.1.
Hyperons were produced by 450~\gevc1 protons impinging on a 40~cm long
beryllium target with a diameter of 0.2~cm.
A magnetic channel consisting of 3 magnets 
selected negative particles 
with a mean momentum of 345~\gevc1, a momentum spread of $\sigma (p)/p=9\%$,
and an angle to the proton beam smaller than 0.5~mrad.
After travelling a distance of 16~m the secondary beam hit the
charm production target which consisted of one 4~mm thick copper plate 
followed by three 2~mm thick carbon plates made of diamond powder with a density of
3.3~g/cm$^3$. The target plates were spaced by 2~cm.
At the target the beam had a width of 3~cm and 
a height of 1.7~cm. Its angular dispersion was 0.6~mrad in the horizontal plane and
1.0~mrad in the vertical plane.
An incoming intensity of $4.0\cdot 10^{10}$ protons per 2.1 second spill
yielded about 
$1.8\cdot 10^{5}$ \sigmamin\ hyperons and $4.5\cdot 10^{5}$ \pimin\ at the
charm production target. 
A transition radiation
detector (TRD) \cite{BRU96} was used to discriminate online between 
\pimin\ and hyperons. 

\begin{figure}[ht]
 \scalebox{0.55}{
   \rotatebox{270}{
      \includegraphics{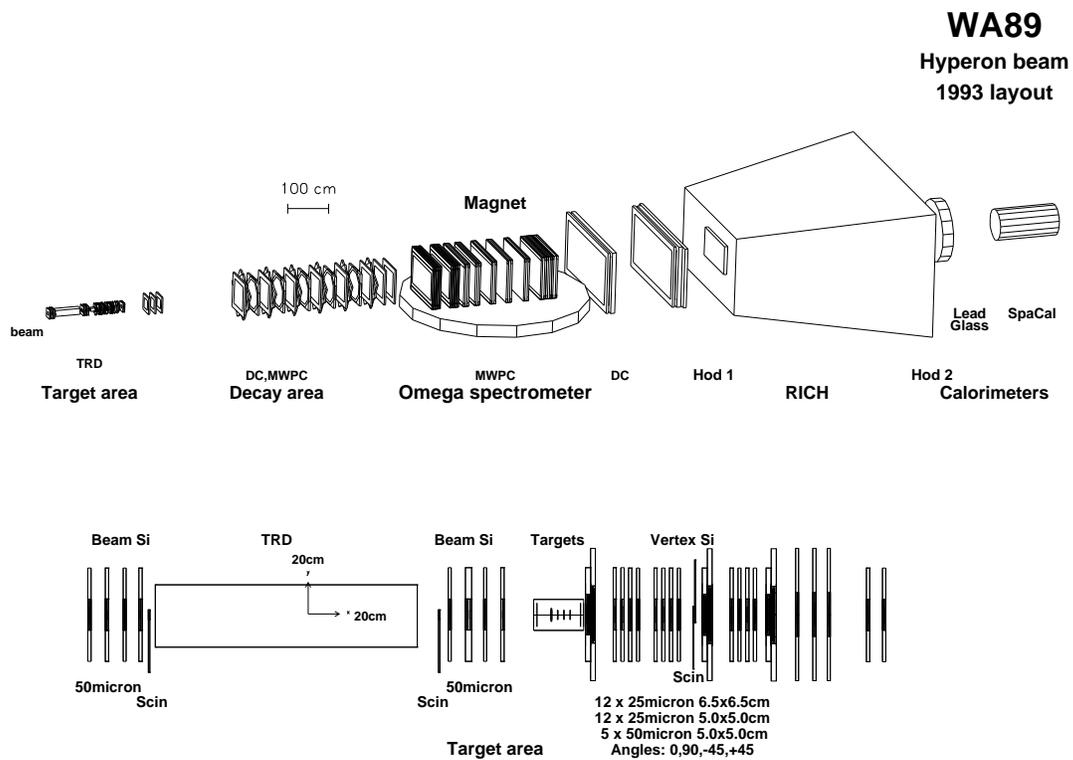}
   }
  }

\caption{Setup of the WA89 experiment in the 1993 run.
        The lower part shows an expanded view of the target area.}
\label{fig:setup}

\end{figure}

Fig.~\ref{fig:setup} shows the experimental setup.
The incoming beam particles and the secondary particles were detected
by a set of silicon micro-strip planes with 25 and 50$\mu $m pitch.
The micro-strip detectors measured the horizontal, vertical and 
$\pm{}$45$^{o}$ projections. 

Positioning the charm production target about 14~m
upstream of the centre of the $\Omega $-spectrometer provided a 10~m long
decay volume for short living strange particles. 
The charged products
of these decays along with the particles coming directly from the target
were detected by 40 planes of drift chambers with a spatial resolution
of about 300~$\mu $m. Special MWPC chambers (20 planes with 1mm wire spacing)  
were used in the central region of high particle fluxes.
In order to improve the track bridging between the target region and the
decay region three sets of 4 MWPCs each with a pitch of 1 mm
were installed about 2~m downstream of the charm production target.
The particle momenta were measured by the
$\Omega$-spec\-tro\-me\-ter \cite{BEU77} consisting of a super-con\-duc\-ting 
magnet with a field integral of 7.5~Tm and a tracking detector
consisting of 45 MWPC planes inside the field and 12 drift chamber
planes at the exit of the magnet. The momentum resolution was
$\sigma (p)/p^{2}\approx 10^{-4}~(\gevc1)^{-1}$.

Charged particles were identified
using a ring imaging Cherenkov (RICH) detector \cite{RICH}. This had a 
threshold of $\gamma~=42$  and provided 
K/$\pi{}$ separation from about 8 to 90~\gevc1{}
with about 90\%{} efficiency and a pion rejection of a factor 10 or more. 
The geometrical aperture of the RICH was nearly 100\% for particles
with momenta $>$15~\gevc1{}. 
Downstream of the RICH a lead glass electromagnetic  
calorimeter was positioned for photon and electron detection \cite{BRU92}.
This calorimeter was in turn followed by a hadron calorimeter\cite{SCH94}.

The trigger was relatively open.
It selected about 25$\% $ of all 
interactions using the multiplicity measured in scintillator
hodoscopes and proportional chambers and using correlations of hits in these
detectors to select particles with high momenta. At least two charged particles
at the exit of the magnet were required.
The results shown in the present paper are based on 
the analysis of about 300 million events recorded in 1993 and 1994.
%
\section{Data Analysis}
\label{sec_data}
The goal was to compare leading and non--leading charmed 
particles which are charge--conjugate: \Dm{} and \Dp{},
\Dsm{} and \Dsp{} and \Lc{} and \Lca{}. 
We looked for  
3-body decays: 
\Dm{}$\rightarrow{}\mathrm{K}^{+}\pi{}^{-}\pi{}^{-}$,
\Dsm{}$\rightarrow{}\mathrm{K}^{+}\mathrm{K}^{-}\pi{}^{-}$,
\Lc{}$\rightarrow{}\mathrm{K}^{-}\mathrm{p}\pi{}^{+}$,
and the charge--conjugate states. 
The acceptance of the setup
is nearly independent of the particle charge (see \ref{subsec_acceptance}) 
and therefore 
the systematic errors of a comparison of conjugate decay modes
should be minimal. 
The key elements of the charm identification are geometrical reconstruction
of events with two or more vertices and identification of kaons
with the RICH.

\subsection{Event reconstruction}
\label{subsec_reconstr}
The reconstruction started with a fast rejection of events which
had no valid interaction in the target by cutting on the
track multiplicities before and after the target.
Additionally, secondary interactions were suppressed by rejecting large
differences in multiplicity between the target region and the
spectrometer.
Charged particle trajectories were reconstructed in the
$\Omega$ spectrometer (for a detailed description see \cite{LASALLE80}),
in the decay region and vertex detector,
then all track segments were bridged together if possible.
Particle identification of protons and kaons was performed with
the RICH. 

The charm search was based on a ``candidate--driven approach''.
All relevant combinations of 3 tracks 
were considered and the coordinates of their origin -- the potential
secondary vertex --
were reconstructed. The rest of the tracks, including the beam
track if it was found, were used to form
the interaction point -- the primary vertex -- in the target. 
These tracks were subjected 
to a filter in order to remove tracks coming from decays
of associated charmed particles, misreconstructed tracks
and other tracks which did not originate in the interaction 
point. 
Some of such tracks passed through the filter causing
systematic distortions of the primary vertex. 
The most powerful  
charm selection criterion was rejection of events with a small
separation between the secondary and primary vertices.
The typical resolution on the longitudinal distance between
these vertices was about 0.5~mm while the typical mean decay lengths of
the charmed particles we looked for were 3--15~mm.
One set of selection criteria was used for all charm
decays considered, including the following cuts:
\begin{list}{--}
  \item a longitudinal separation between the vertices of 
        $>$8--9$\sigma{}$, depending on the particle type;
  \item the secondary vertex to be outside of the target material by 
        $>1-3\sigma{}$;
  \item impact parameters of the tracks from the secondary vertex to 
        the primary vertex of 
        $>$2$\sigma{}$;
  \item an impact parameter of the reconstructed track of the charm particle candidate to 
        the primary vertex of 
        $<$4$\sigma{}$;
  \item ``soft'' RICH identification of kaons and protons, mainly 
        rejecting clearly identified pions.
\end{list}

The values of the cuts were identical  for the charge--conjugate modes, while 
the above cuts on vertex separation were slightly different for 
different particles because of the different decay geometries.

The resulting invariant mass spectra for the six decays modes are
presented in Fig.~\ref{fig:massplots}. Clear signals are observed
for \Dm{}, \Dp{}, \Dsm{} and \Lc{}, while no significant
mass peaks are observed for \Dsp{} and \Lca{}. Small signals 
of  about 5-10 events for
the two latter particles are observed with harder 
cuts. Such harder cuts, however, reduce the signals
for \Dsm{} and \Lc{} and increase the statistical error on the
estimate of the production asymmetry. The numbers of observed 
signal and background events as well
as the deviations of the signals from the PDG~\cite{PDG96} masses
are shown in Table \ref{tab:particles}. 
There are no multiple entries from one event to the signal regions.
Possible mutual reflections
of the mass peaks have been studied. Due to a good kaon and proton
identification these reflections are not strong. 
The contamination of the \Dsm{} signal
with \Dm{} is estimated to be below 8 events, and  
the contamination of the \Lc{} signal
with \Dp{} is estimated to be below 3 events.

\subsection{Acceptance Corrections}
\label{subsec_acceptance}
%
If the acceptance of the experiment is truly charge independent
one could measure the \xf{} dependence of the
asymmetry between two charge--conjugate states
without correcting for the acceptance,
provided that the bins in \xf{} were small enough.
In WA89 the acceptance was
nearly symmetric, however the efficiency of the trigger and the RICH 
introduced a slight charge asymmetry. 
Therefore we used the full 
calculation of charm reconstruction efficiencies
in order to correct the \xf{} spectra observed. 
Charm events were simulated by an inclusive
event generator that generated the
charm particle considered, according to the measured kinematic distributions
together with an associated anti--charm particle.
The latter was produced with the same kinematic distributions
as the first particle, taking into account
an azimuthal angular correlation
with the first particle, measured in a pion beam~\cite{WA9296}.
For the associated particles a mixture of about 
25\%{} of charged D-mesons, 
50\%{} of neutral D-mesons, and 
25\%{} of \Lc{} was taken, in accordance with average yields of charmed
particles measured in various experiments.  
The remaining multiplicity of the interaction was generated using the 
FRITIOF Monte Carlo for hadron--nucleus interactions \cite{FRITIOF}.
The particles were tracked through the detector by a detector
simulation program based on the GEANT package \cite{GEANT321}. The response functions
of the different detector components and the trigger were simulated according to
the measured efficiencies. The trigger logic was simulated as well.
The simulated events were then analysed using the regular track reconstruction and
charm reconstruction procedure, and the acceptances of the experiment
to the decay modes studied were evaluated applying the whole set of
selection criteria used for the data. The number of reconstructed
simulated charm events was at least one order of magnitude higher
than the size of the charm signals observed in the data.

\begin{figure}[ht]
 \begin{center}
 \scalebox{1.4}{
   \rotatebox{270}{
      \includegraphics{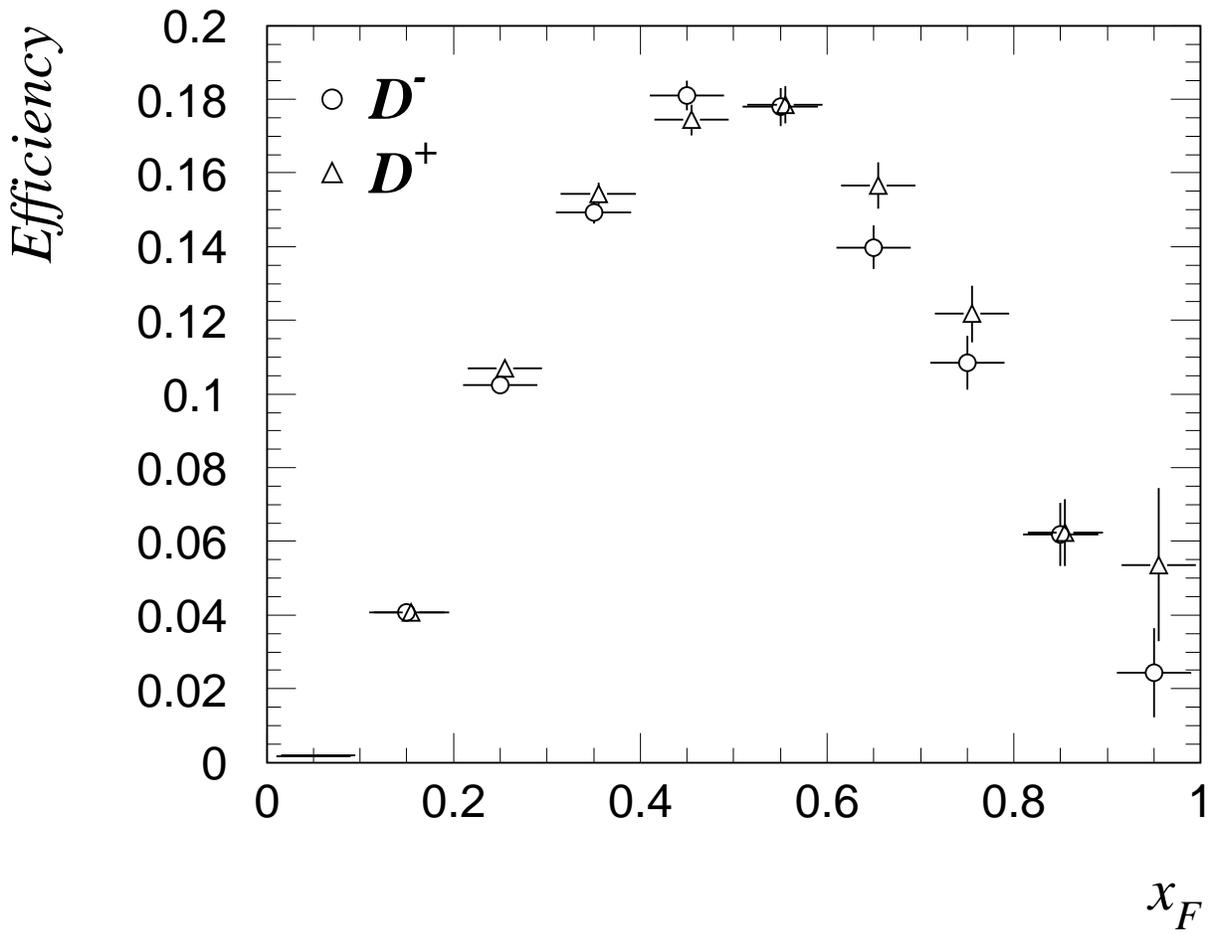}
   }
  }
\end{center}

\caption{Full efficiency of D$^\pm$ meson reconstruction. The average values
         for the 1993 and 1994 runs of data taking are shown.
        }
\label{fig:acceptance}
\end{figure}

The acceptance of the experiment 
does not depend strongly on the transverse momentum
of the charmed particles up to \pt{}=2~\gevc1{}. Most of 
the charm events
observed had \pt{}$<$1.5~\gevc1{}. 
Therefore only the \xf{} dependence of the acceptance was considered.
These dependences for \Dm{} and \Dp{} averaged for the 1993 and 1994
runs of data--taking  are shown in Fig.~\ref{fig:acceptance}. 
The acceptances
drop sharply at \xf{}$<$0.2 and are very low at \xf{}$<$0.1.
A loss of efficiency at high \xf{} is caused by the charm reconstruction
technique which required the reconstruction of the primary 
vertex built of leftover tracks. 
For high \xf{} there are fewer leftover particles  
with momentum high enough
to be reconstructed.
The efficiencies for the charge--conjugate modes differ by about 10\%. 
The calculated charm reconstruction efficiency for the 1994 run was about 
40\% higher than for 1993 (see Table \ref{tab:rawdat}),
due to a better vertex detector and the choice of charm selection cuts. 
However for this analysis only the relative efficiencies for charge
conjugated modes within one run are important.

\subsection{Production Asymmetries}
\label{subsec_asymmetries}
%
Comparison of the observed raw number of events shows 
a considerable enhancement of the leading over non--leading
particles in the useful kinematic range of \xf{}$>$0.1, 
taking into account that the acceptances for them
are close. For a quantitative estimate
the data were split in bins of \xf{} separately for the 1993 and 1994 data. 
The value of \xf{} was 
calculated using a fixed beam momentum of 340~GeV/c.
Since the number of events observed in a bin
was low, in particular at high \xf{},
we calculated the values of the asymmetry and the errors using
the maximum likelihood method.
For a given \xf{} bin we took the number of events $N$ -- in the signal
mass window of $\pm{}$20~\Mevc2{} and $B$ -- in the sidebands of 
-140 to -20~\Mevc2{} and 20-140~\Mevc2{} with respect to the table 
mass of the particle. 
The numbers $N$ and $B$ along with the calculated efficiencies $\cal E$
and its errors
are summarised in Table \ref{tab:rawdat}.
Since the background in the mass spectra
is well described by a linear distribution, we estimate
the expected number of background events in the signal window
as $\alpha{}\cdot{}B$, where $\alpha{}=1/6$. 
The parameters used were the expectation values for the background in the sidebands 
$b_{1,2}$, the expectation values for the efficiencies $\epsilon{}_{1,2}$, 
where the indices 1,2 refer to the leading and non-leading particles, respectively,
and the expectation value for the combined signal $s$. The expectation values
for the leading and non-leading signals were obtained for a given asymmetry $A$ as:
$s_{1,2}~=~s\cdot{}(1~\pm{}~A)\cdot{}\epsilon_{1,2}$, where $\pm$ sign
refer to the leading and non--leading particles, respectively.
The likelihood function then was:

\begin{equation}
{ \cal L}(A) = \prod_{i=1,2}
  \frac{e^{-(s_{i}+\alpha{}b_{i})}\cdot{}(s_{i}+\alpha{}b_{i})^{N_{i}}}{N_{i}!}
       \times{}
   \frac{e^{-b_{i}}\cdot{}b_{i}^{B_{i}}}{B_{i}!} \times{} e^{-((\epsilon_{i}-{\cal E}_{i})/\sigma_{i})^{2}/2} ,
\label{eq:ps}
\end{equation}

where the last term takes into account the statistical errors of the efficiency
calculations.
The likelihood function ${\cal L}(A)$ was maximised   
for each value of $A$
with respect to
the other parameters, taking into account the physical limits
$s,b_{1,2}\geq{}0$.
The logarithms of the likelihood functions for the 1993 and 1994 data were added
and the expectation value of asymmetry
$\overline{A}$ was found at the maximum of the likelihood function ${\cal L}_{max}$.
The 1$\sigma$ confidence interval was found at the points of 
$ln({\cal L}_{max})-ln({\cal L})=0.5$ 
within the allowed range $-1\leq{}A\leq{}1$. If only one such point existed 
a one-side error is given.  
Lower limits on the asymmetry at 90\% confidence level, $A_{min}$, were obtained
looking for a value $A < \overline{A}$ at
$ln({\cal L}_{max})-ln({\cal L})=\Delta{}$, with $\Delta{}$ varying from $\Delta$=0.82
for $\overline{A}$ being at least 2$\sigma$ away from $\pm{}1$ to $\Delta$=1.35
for $\overline{A}$=1.

The results are presented in Table \ref{tab:asym} and in Fig.~\ref{fig:asym}.
Only the statistical errors are shown.

\subsection{Systematic Errors}
\label{subsec_systematics}
Systematic errors may come from charge asymmetries in the detection efficiencies,
not accounted for in the simulation, or from beam contamination by particles
different from \sigmamin{}. 
The spectrometer acceptance does not impose any considerable charge asymmetry,
though the acceptance of the RICH was slightly asymmetric. Possible effects
of this asymmetry were studied comparing accepted and rejected events in
different momentum ranges. 
Kaon identification criteria were not stringent. A typical efficiency of
about 90\%{} for 
such criteria was measured 
using clearly identifiable $\phi{}\rightarrow$K$^+$K$^-$ decays. 
On the other
hand, if stronger vertex cuts were used, relatively clean charm signals
were seen even with very weak RICH cuts. 
It was observed that indeed 
the efficiency of the RICH cuts was not worse than 90\%{}. A conservative
estimate of the systematic error of $A$ associated with the RICH efficiency 
is 3\%{} over the full \xf{} range.
 Another possible source of systematic errors is the efficiency of the trigger,
whose logic was charge dependent. In simulation the trigger caused 
an asymmetry of about 10\%{} at high momenta. The trigger simulation
has been compared with the data using a sample of minimum bias events recorded
with minimal trigger requirements. Within these data a sample of \lambdazero{}
decays was found, its trigger efficiency measured and compared
to simulation of \lambdazero{} production. The simulated trigger efficiency was about
0.85~--~0.95 of the measured one. The difference can be attributed mainly to
the multiplicity in the last scintillator hodoscope being higher in data than
in simulation and is not charge dependent.
A systematic error on $A$ 
of 5\%{} is assigned to account for uncertainties of the trigger efficiency.

Stability of the results on variations of the most important geometrical
cuts was also studied. No statistically significant change of the measured
asymmetries has been observed.  

Another type of systematic error comes from beam contamination.
The beam was identified by the TRD used in the trigger. During the data taking
a sample of events with no beam interaction requirement was recorded.
This sample was used to study the beam composition.
\sigmamin{} could be identified by their decays.
The main 
contamination was \pimin{} comprising 11\%{} and 18\%{} of the
events used for the charm search for the 1993 and 1994 data respectively.
This contamination could be reduced by an off-line analysis of the TRD
information. Unfortunately,
the full TRD information was available only for a part of the selected 
charm sample.
This subsample was used to estimate the impact of the pion contamination.
This impact depends on the type of the charmed particle studied.
It was observed that the \Dpm{} signal splits
between the \sigmamin{} and \pimin{} beam subsamples proportionally
to the beam composition,
within the statistical accuracy, but no statistically significant
\Ds{} or \Lc{} signal was observed in the pion beam subsample.
\asymd{} measured in a \pimin{} beam at 500~GeV/c \cite{E79196} is
shown in Fig.~\ref{fig:asym} as a curve fitted to these data.
The results of our measurement 
indicate slightly higher asymmetries.
If \asymd{} in both \sigmamin{} and \pimin{} beams were the same
no systematic error would be added by a \pimin{} contamination. 
In the extreme assumption
that \asymd{} in a \sigmamin{} beam were two times higher than in a \pimin{}
beam, the contamination may dilute the measured asymmetry by a factor of about
0.92. For the \Ds{} and \Lc{}, not more than 10\%{} of the sample may originate from 
the pion component of the beam. Since no asymmetry for these particles has been observed 
in pion beams, a possible dilution factor in our measurement is not lower than 0.90.
Another source of beam contamination comes from the decays of \sigmamin{} in the beam 
to \pimin{} and neutrons, when one of the decay products interacts in the charm
production target. 
The pions cannot produce a competitive sample of charmed particles
due to their low energy, but the neutrons can. In this data analysis
no requirement of a beam track association with the reconstructed
primary vertex was applied and therefore some part of the charm sample
may be produced by neutrons. A number of neutron interactions at the 5\%{} level
was observed
in the full sample of interactions recorded. 
Since neutrons contain two d-quarks, similar to \sigmamin{}, one may expect
similar asymmetries for \Dpm{} and \Lc{} produced by
neutrons and \sigmamin{} and therefore no dilution for \asymd{} and \asymlc{}.   
If \Dsm{} production by neutrons were comparable with \Dsm{} production by \sigmamin{}
\asymds{} could at most be diluted 
by a factor of about 0.95.
The \Dsm{} signal is contaminated with not more than 8 \Dm{} events.  
If \asymds{}=\asymd{} 
no systematic error would be added to \asymds{} by this contamination. 
In the extreme assumption that
\asymds{}=2\asymd{} 
the contamination may dilute the measured asymmetry by a factor of about
0.95. The contamination of 
the \Lc{} signal
with \Dp{} is below 3 events and its role is less significant.

For the estimate of the full systematic error of the asymmetry measurement
we added in quadrature the errors
coming from the efficiency uncertainties and obtained a value of about 6\%{}
over the full \xf{} range and for all the particles considered.
Additionally, the values of the observed asymmetries may be diluted by a
a factor of $>$0.9 due to the beam contamination. These systematic
errors are smaller than the statistical errors of the measurement and therefore
they are not considered in Table \ref{tab:asym} and in Fig.~\ref{fig:asym}.

\section{Discussion}
\label{sec_discussion}

\begin{figure}[ht]
 \begin{center}
 \scalebox{0.7}{
      \includegraphics{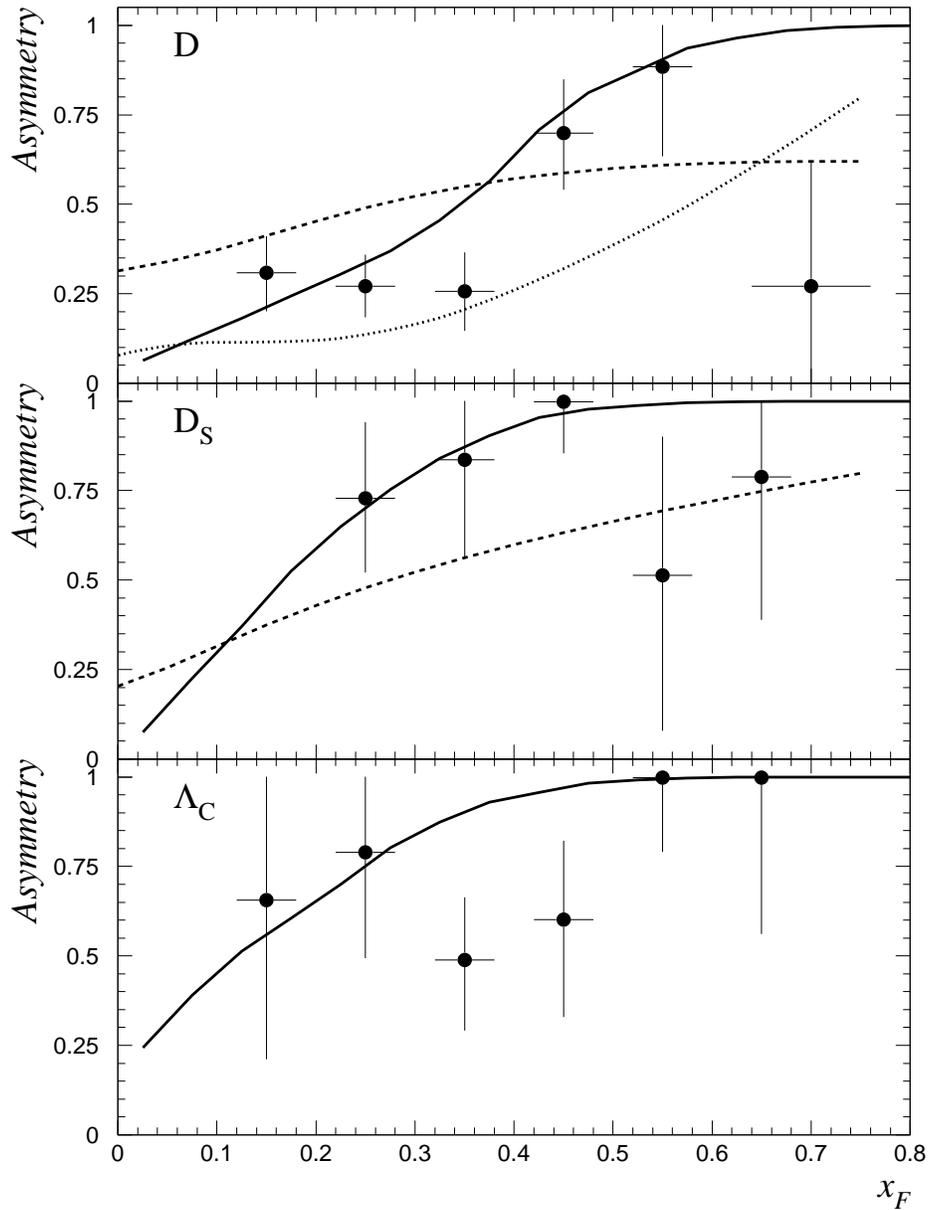}
  }
 \end{center}

\caption{Asymmetries for \Dpm{}, \Ds{}, and \Lc{} production. The solid and
         dashed curves show the results of calculations \cite{SLABOSP97p} and 
         \cite{PISKOUN97p}, respectively.
         The dotted curve shows the tuned PYTHIA calculations for \pimin{} beam 
         taken from \cite{E79196},  which represent closely the results obtained by 
         experiment E791 at FNAL. 
        }
\label{fig:asym}

\end{figure}

The asymmetry \asymd{} observed (Fig.~\ref{fig:asym})
is similar 
to the results of pion beam measurements, though it is generally higher 
than the pion beam data by 1-2 standard deviations. The drop
of the last point at \xf{}=0.7 is not statistically significant.
The asymmetries \asymds{} and \asymlc{} are generally higher than \asymd{}.

The results have been compared with two theoretical calculations.
In a parton model approach applied in \cite{SLABOSP97} and earlier papers 
\cite{SLABOSP80s}
PQCD is used to describe
the charm quark production. The model considers two mechanisms contributing
to the subsequent hadronisation.
The first one is fragmentation
describing charm quark hadronisation in the absence 
of any interaction with
quarks from initial hadrons.
This process has been studied extensively in $e^{+}e^{-}$ experiments
and has been well described phenomenologically in terms of fragmentation
functions. The second mechanism is recombination with valence quarks.
A recombination function was constructed by analogy with the fragmentation
functions and certain boundary conditions were applied to it. 
The resulting model depends on a few parameters included in the recombination function
which were tuned to the charm production asymmetries
measured in pion beams \cite{SLABOSP97}.
The main uncertainty comes from an unknown relative contribution of
the recombination process.
Using the same set of parameters
predictions for the \sigmamin{} beam were obtained\cite{SLABOSP97p}. 
The resulting predictions 
are shown
in Fig.~\ref{fig:asym} and describe the data well.

Another theoretical calculation was performed applying the
quark--gluon--string model (QGSM) to charm particle production
\cite{PISKOUN97,PISKOUN97p,ARAKEL97} where charm particles are created in the break up of strings
linked to valence quarks in the incoming hadron.
In this approach a strong leading/non--leading
production asymmetry at high \xf{} emerges from the momentum correlation of
charm and valence quarks.
This model was extended by
adding a certain amount of charm sea similar to the one suggested in 
the ``intrinsic charm'' model \cite{BRODSKY96}.
The charm quarks and anti-quarks from the ``charm sea''
are assumed to have similar ``hard'' momentum distributions. 
In the string breaking process they are treated similarly to the valence quarks and
the resulting production of charmed hadrons from the ``charm sea''
does not involve 
``normal'' valence quarks of the colliding hadrons.
Therefore the ``charm sea'' in this model 
contributes equally to the leading and
non-leading particles at high momenta thus diluting the
asymmetry
\footnote{It should be pointed out that the ``intrinsic
  charm''   model\cite{BRODSKY96} 
  led to a different result. 
  This model considered recombination of the ``intrinsic'' charm quarks  
  with valence quarks as an important process because of their close velocities. 
  Therefore 
  the valence quarks contribute to the momenta of the leading particles 
  making them ``harder'' than the non--leading ones. The model predicted
  a considerable asymmetry \asymd{} at \xf{}$>$0.5 for \pimin{} beams, however
  this prediction did not fit well to the data\cite{E79196}.
  }.
The model contains two parameters one of which is the intrinsic charm amount,
and should describe the production asymmetries
and full charm cross--sections. 
The parameters have been tuned to the charm production asymmetries 
measured in pion beams. Using the same set of parameters
predictions for \asymd{} and \asymds{} in 
\sigmamin{} beam were obtained\cite{PISKOUN97p}.  
These predictions
are shown
in Fig.~\ref{fig:asym} and also describe the data reasonably well.

\section{Summary and conclusions}

We have presented the first measurement of charge asymmetries
of charmed particles produced in a \sigmamin{} beam. Significant asymmetries
in favour of leading over non--leading particles have been observed.
The asymmetry observed for \Dpm{} production is compatible with the results 
obtained in pion beams. The results on \Ds{} production point
to a strong leading particle effect involving strangeness from the beam.
And finally, a strong production asymmetry in favour of \Lc{} baryon 
over \Lca{} anti-baryon is observed, in contrast with \Lc{} production
by pions. This confirms the indication of a strong leading
particle effect with charmed baryons produced by baryon beams,
reported by experiment E769\cite{E769a96}. 

\section*{ Acknowledgements }

We are indebted to J.~Zimmer and the late Z.~Kenesei
for their help during all moments of detector construction and set-up.
We are grateful to the staff of CERN's 
EBS group for providing an excellent
hyperon beam channel, to the staff of CERN's Omega group for their 
help in running
the $\Omega $-spectrometer and also to the staff of the SPS for providing
good beam conditions. 
We thank 
O.~Piskounova and S.~Slabospitsky for providing the predictions of their
models for this experiment and 
G.H.~Arakelyan and A.K.~Likhoded for helpful discussions.
The collaboration with O.~Piskounova was supported 
by DFG (Germany) and RFBR (Russia) under contract number 436 RUS 113/332/2.

\begin{table}[ht]
\begin{center}

\begin{tabular}{|l|l|c|c|c|} \hline
              &        & \# signal      & Deviation from            &  \\
  Particle    & Decay  & and background & the PDG \cite{PDG96}      & Width \Mevc2  \\
              &        &                & mass \Mevc2{}             &  \\ 
 \hline
   \Dm{}   & $\mathrm{K}^{+}\pi{}^{-}\pi{}^{-}$       &  296/38 &  0.3$\pm{}$1.5 & 8.5$\pm{}$0.5  \\
   \Dp{}   & $\mathrm{K}^{-}\pi{}^{+}\pi{}^{+}$       &  142/42 & -0.7$\pm{}$1.6 & 5.9$\pm{}$0.5    \\
 \hline
   \Dsm{}  & $\mathrm{K}^{+}\mathrm{K}^{-}\pi{}^{-}$  &   76/24 &  0.7$\pm{}$1.7 & 4.5$\pm{}$0.8    \\
   \Dsp{}  & $\mathrm{K}^{+}\mathrm{K}^{-}\pi{}^{+}$  & $<$27 90\% CL &  &  \\
 \hline
   \Lc{}   & $\mathrm{K}^{-}\mathrm{p}\pi{}^{+}$      &   74/20 &  1.9$\pm{}$1.7 & 5.0$\pm{}$0.6    \\
   \Lca{}  & $\mathrm{K}^{+}\overline{\mathrm{p}}\pi{}^{-}$  & $<$24 90\% CL &  & \\
 \hline
\end{tabular}

\end{center}
\caption{ 
  Number of reconstructed signal events, background events  
  and parameters 
  of the mass fit.
}
\label{tab:particles}
\end{table}

\begin{table}[htb]
\begin{center}

\begin{tabular}{|c||r|r|c||r|r|c||r|r|c||r|r|c|} \hline
      & \multicolumn{6}{c||}{\Dm{}} 
      & \multicolumn{6}{c|}{\Dp{}} 
  \\ \cline{2-13}
        \xf{}
      & \multicolumn{3}{c||}{run $\#$1} 
      & \multicolumn{3}{c||}{run $\#$2} 
      & \multicolumn{3}{c||}{run $\#$1} 
      & \multicolumn{3}{c|}{run $\#$2} 
  \\ \cline{2-13}
      & $N$ & B & $\cal E$ \% & $N$ & B & $\cal E$ \% 
      & $N$ & B & $\cal E$ \% & $N$ & B & $\cal E$ \%
  \\ \hline
      0.1 - 0.2
       & 23 & 17 &  3.1$\pm$0.1  & 51 & 29 &  5.1$\pm$0.2
       & 15 & 11 &  3.3$\pm$0.1  & 26 & 25 &  4.9$\pm$0.2
  \\
      0.2 - 0.3
       & 44 & 36 &  8.4$\pm$0.3  & 62 & 36 & 12.1$\pm$0.3
       & 28 & 37 &  8.8$\pm$0.3  & 40 & 34 & 12.6$\pm$0.3
  \\
      0.3 - 0.4
       & 31 & 19 & 12.3$\pm$0.4  & 41 & 41 & 17.6$\pm$0.4
       & 27 & 23 & 12.8$\pm$0.4  & 19 & 28 & 18.1$\pm$0.4
  \\
      0.4 - 0.5
       & 20 & 24 & 14.4$\pm$0.6  & 28 & 19 & 21.8$\pm$0.6
       &  8 & 26 & 14.7$\pm$0.6  &  7 & 21 & 20.2$\pm$0.6
  \\
      0.5 - 0.6
       &  8 &  5 & 14.5$\pm$0.7  & 13 &  7 & 21.1$\pm$0.8
       &  0 &  7 & 15.5$\pm$0.7  &  5 & 13 & 20.2$\pm$0.7
  \\
      0.6 - 0.7
       &  5 &  3 & 10.4$\pm$0.9  &  2 &  3 & 14.5$\pm$1.0
       &  4 &  6 & 11.5$\pm$0.9  &  2 & 11 & 16.3$\pm$1.0
  \\
      0.7 - 0.8
       &  2 &  1 &  7.4$\pm$0.9  &  2 &  1 & 14.3$\pm$1.1
       &  1 &  0 &  9.8$\pm$1.0  &  0 &  3 & 14.5$\pm$1.1
  \\ \hline \hline
      & \multicolumn{6}{c||}{\Dsm{}} 
      & \multicolumn{6}{c|}{\Dsp{}} 
  \\ \hline
      0.1 - 0.2
       &  5 & 17 &  2.2$\pm$0.1  &  8 & 35 &  3.6$\pm$0.1
       &  6 & 17 &  2.5$\pm$0.1  &  6 & 24 &  3.2$\pm$0.1
  \\
      0.2 - 0.3
       & 13 & 19 &  6.5$\pm$0.2  & 25 & 19 &  9.5$\pm$0.2
       & 13 & 30 &  6.9$\pm$0.2  &  4 & 23 &  9.9$\pm$0.2
  \\
      0.3 - 0.4
       &  7 & 12 &  9.4$\pm$0.3  & 15 & 27 & 13.5$\pm$0.3
       &  4 & 21 & 10.7$\pm$0.3  &  3 & 12 & 14.4$\pm$0.3
  \\
      0.4 - 0.5
       &  6 &  9 & 12.0$\pm$0.5  & 10 &  8 & 15.5$\pm$0.4
       &  3 &  9 & 12.7$\pm$0.6  &  0 & 10 & 17.8$\pm$0.4
  \\
      0.5 - 0.6
       &  2 &  5 & 11.6$\pm$0.8  &  5 &  5 & 16.6$\pm$0.6
       &  1 &  2 & 12.3$\pm$0.9  &  2 &  5 & 17.4$\pm$0.6
  \\
      0.6 - 0.7
       &  4 &  0 &  8.4$\pm$1.3  &  0 &  1 & 15.4$\pm$0.7
       &  1 &  2 & 11.8$\pm$1.5  &  2 &  2 & 16.6$\pm$0.8
  \\
     \hline \hline
      & \multicolumn{6}{c||}{\Lc{}} 
      & \multicolumn{6}{c|}{\Lca{}}
  \\ \hline
      0.1 - 0.2
       &  7 & 16 &  1.0$\pm$0.1  & 10 & 20 &  1.3$\pm$0.1
       &  3 & 23 &  0.9$\pm$0.1  &  7 & 27 &  1.2$\pm$0.1
  \\
      0.2 - 0.3
       &  9 & 16 &  2.5$\pm$0.1  & 13 & 12 &  3.6$\pm$0.2
       &  3 &  9 &  2.3$\pm$0.1  &  3 & 18 &  2.8$\pm$0.2
  \\
      0.3 - 0.4
       & 14 &  7 &  4.2$\pm$0.2  & 12 & 11 &  5.5$\pm$0.3
       &  1 &  5 &  3.3$\pm$0.2  &  6 &  2 &  4.3$\pm$0.2
  \\
      0.4 - 0.5
       &  4 &  4 &  4.5$\pm$0.3  &  9 &  4 &  6.3$\pm$0.4
       &  0 &  1 &  4.4$\pm$0.4  &  3 &  2 &  5.4$\pm$0.3
  \\
      0.5 - 0.6
       &  0 &  2 &  4.4$\pm$0.3  &  5 &  2 &  6.1$\pm$0.4
       &  0 &  0 &  3.6$\pm$0.6  &  0 &  0 &  6.1$\pm$0.4
  \\
      0.6 - 0.7
       &  0 &  3 &  3.2$\pm$0.4  &  3 &  3 &  6.2$\pm$0.6
       &  0 &  0 &  2.8$\pm$0.9  &  0 &  0 &  5.1$\pm$0.5
  \\
     \hline 
\end{tabular}

\end{center}
\caption{The numbers of events observed in the signal mass band $N$,
         the number of background events observed in the side bands $B$  
         and the calculated reconstruction efficiencies.  
         The background expected in the signal band is about 16.7\% of the observed
         one. 
         }
\label{tab:rawdat}
\end{table}

\renewcommand{\arraystretch}{1.3}

\begin{table}[htb]
\begin{center}

\begin{tabular}{|c||c|r||c|r||c|r|} \hline
    & \multicolumn{2}{c||}{\Dpm{}} 
    & \multicolumn{2}{c||}{\Ds{}} 
    & \multicolumn{2}{c|}{\Lc{}} 
  \\ \cline{2-7}
      \multicolumn{1}{|c||}{\xf{}}
    & $\overline{A}$ & $A_{min}$ 
    & $\overline{A}$ & $A_{min}$ 
    & $\overline{A}$ & $A_{min}$ 
  \\ \hline
 0.1 - 0.2
              &    0.31$^{+0.10}_{-0.11}$  &   0.17
              &                            &  
              &    0.66$^{+    }_{-0.44}$  &   0.02
\\
 0.2 - 0.3
              &    0.27$^{+0.09}_{-0.09}$  &   0.16
              &    0..73$^{+0.21}_{-0.21}$  &   0.45
              &    0.79$^{+    }_{-0.29}$  &   0.37
\\
 0.3 - 0.4
              &    0.26$^{+0.11}_{-0.11}$  &   0.12
              &    0.84$^{+    }_{-0.27}$  &   0.43
              &    0.49$^{+0.17}_{-0.20}$  &   0.23
\\
 0.4 - 0.5
              &    0.70$^{+0.15}_{-0.16}$  &   0.49
              &    1.00$^{+    }_{-0.14}$  &   0.66
              &    0.60$^{+0.22}_{-0.27}$  &   0.24
\\
 0.5 - 0.6
              &    0.88$^{+    }_{-0.25}$  &   0.51
              &    0.51$^{+0.39}_{-0.44}$  &  -0.08
              &    1.00$^{+    }_{-0.21}$  &   0.48
\\
 0.6 - 0.7
              &                            &  
              &    0.79$^{+    }_{-0.40}$  &   0.10
              &    1.00$^{+    }_{-0.44}$  &   0.00
\\
 0.6 - 0.9
              &    0.27$^{+0.34}_{-0.37}$  &  -0.20
              &                            &  
              &                            &  
\\
 \hline
\end{tabular}

\end{center}
\caption{ The expectation values $\overline{A}$ and lower limits at 90\% CL $A_{min}$
          for the asymmetries measured.
          Only the statistical errors are shown.}
\label{tab:asym}
\end{table}

\renewcommand{\arraystretch}{1.}

\begin{figure}[ht]
\begin{center}
\mbox{\epsfxsize=12cm\epsffile{Fig2.epsi}}
\end{center}

\caption{Invariant mass distributions for 
        {\bf (a)} \Dm{}$\rightarrow{}$K$^{+}\pi{}^{-}\pi{}^{-}$,
        {\bf (b)} \Dp{}$\rightarrow{}$K$^{-}\pi{}^{+}\pi{}^{+}$,
        {\bf (c)} \Dsm{}$\rightarrow{}$K$^{+}\mathrm{K}^{-}\pi{}^{-}$,
        {\bf (d)} \Dsp{}$\rightarrow{}$K$^{-}\mathrm{K}^{+}\pi{}^{+}$,
        {\bf (e)}  \Lc{}$\rightarrow{}$K$^{-}\mathrm{p}\pi{}^{+}$,
        {\bf (f)} \Lca{}$\rightarrow{}$K$^{+}\overline{\mathrm{p}}\pi{}^{-}$.
        }
\label{fig:massplots}

\end{figure}

\end{document}